\documentclass[conference]{IEEEtran}
\IEEEoverridecommandlockouts
\usepackage{cite}
\usepackage{caption}
\usepackage{amsmath,amssymb,amsfonts}
\usepackage{float}
\usepackage{algorithm}
\usepackage{algorithmic}
\usepackage[top=0.7in, bottom=1in, left=0.65in, right=0.6in]{geometry}

\renewcommand{\baselinestretch}{0.95}

\usepackage{graphicx}
\usepackage{textcomp}
\usepackage{xcolor}
\def\BibTeX{{\rm B\kern-.05em{\sc i\kern-.025em b}\kern-.08em
    T\kern-.1667em\lower.7ex\hbox{E}\kern-.125emX}}

\usepackage[hidelinks]{hyperref}

\begin{document}

\title{Accelerating Development in UAV Network Digital Twins with a Flexible Simulation Framework}

\author{
\IEEEauthorblockN{Md Sharif Hossen, Anil Gurses, Mihail Sichitiu, İsmail Güvenç}  
\IEEEauthorblockA{
Department of Electrical and Computer Engineering, North Carolina State University, Raleigh, USA\\  
Email: \{mhossen, agurses, mlsichit, iguvenc\}@ncsu.edu  
}  \thanks{This work is supported in part by the NSF award CNS-1939334.}
}

\maketitle

\begin{abstract}
Unmanned aerial vehicles (UAVs) enhance coverage and provide flexible deployment in 5G and next-generation wireless networks. The performance of such wireless networks can be improved by developing new navigation and wireless adaptation approaches in digital twins (DTs). However, challenges such as complex propagation conditions and hardware complexities in real-world scenarios introduce a realism gap with the DTs. Moreover, while using real-time full-stack protocols in DTs enables subsequent deployment and testing in a real-world environment, development in DTs requires high computational complexity and involves a long development time. In this paper, to accelerate the development cycle, we develop a measurement-calibrated Matlab-based simulation framework to replicate performance in a full-stack UAV wireless network DT.
In particular, we use the DT from the NSF AERPAW platform and compare its reports with those generated by our developed simulation framework in wireless networks with similar settings. In both environments, we observe comparable results in terms of RSRP measurement, hence motivating iterative use of the developed simulation environment with the DT. 
\end{abstract}


\section{Introduction}
In the last few decades, there has been extensive technological advancement in wireless communication. Mobility, manoeuvrability, low cost, and line of sight (LOS) communication make unmanned aerial vehicles (UAVs) an attractive solution for many real-world challenges in 5G and next-generation wireless networks \cite{in_1}. While UAVs provide many benefits like enhanced coverage and flexible deployment options, there exist several challenges such as complex propagation conditions, operational complexities, and hardware limitations which affect the performance of UAV communications. Moreover, interference can affect reliable communication.

To tackle these challenges, digital twins (DTs) represent a viable solution to replicate and evaluate UAV operations in real-time environments. DT is a virtual replica of the physical system where we can do the emulation of UAV operations without facing the constraints of real-world experimentation. It offers several benefits such as real-time emulation, predictive maintenance, and performance optimization. DTs enable safe testing and optimization of UAV flight paths, control algorithms, and mission planning in a virtual environment by reducing the need for costly real-world test flights.  It also enables researchers to evaluate UAV operations in various scenarios without physical risks and fosters collaborative research for UAV development. 

However, though DTs offer enhanced realism for UAV operations in a virtual environment, there are several challenges in accurately replicating complex real-world factors such as signal propagation conditions and dynamic operational situations. It requires high computational needs involved in creating and maintaining high-fidelity DTs that can handle real-time simulations with full-stack protocols. Hence, we need a joint simulation-emulation framework.
Such a framework can provide a more efficient and cost-effective approach to UAV communication research for testing and validating communication strategies before implementing them in a DT. 

In this paper, we present a measurement-calibrated Matlab-based simulation framework to replicate the performance observed in a DT environment. We use aerial experimentation and research platform for advanced wireless (AERPAW) \cite{ae_1} testbed, which offers rich insights into real-world UAV operations, enabling a thorough evaluation of communication scenarios. We compare the results obtained from our Matlab-based simulation framework with those generated by AERPAW’s DT. We find a high similarity between results from these two environments such as in the reference signal received power (RSRP) levels. Hence, this framework works as a practical tool for researchers to conduct their preliminary studies and simulations for the development of UAV communication strategies in a controlled yet realistic environment. 

The paper is structured as follows. Section~\ref{rw} reviews existing work on UAV simulations and DTs. Section~\ref{smpl} outlines the system model and path loss calculations. An abstraction of the propagation channel and the integration of radio software has been discussed in Section~\ref{rsoft}. Section~\ref{trjlog} includes the discussion of vehicle software and mission plan abstraction. Section~\ref{rd} discusses the results from the comparative analysis of the Matlab simulation and AERPAW emulation. Finally, Section \ref{con} provides concluding remarks and future directions.

\section{Related Works}\label{rw}
Integration of UAV communication in 5G networks has received a growing interest in recent times. While UAVs provide autonomy and coverage benefits, they experience complex challenges in signal propagation, interference, and other communication-related factors such as signal fading and random variations in the communication channel. To analyze the UAV communications performance, simulation frameworks and DTs such as Colosseum \cite{dt} and AERPAW are becoming essential tools that allow researchers to replicate various UAV scenarios and fine-tune wireless and navigation parameters under different deployment conditions \cite{rw_1}.

Various approaches have been proposed to enhance throughput and signal quality to optimize the performance of UAV communication systems with autonomous mobility. For example, the authors in \cite{max_thr} aimed to maximize the throughput among all ground users over a fixed UAV flight time for UAV-enabled wireless-powered communication networks. They proposed a hover-and-fly trajectory with proper wireless resource allocation to maximize throughput. Authors in \cite{lte_alt} proposed an LTE-based system to analyze cellular connectivity with signal strength and other parameters at different altitudes. They showed that by increasing the altitude from ground to 170 m, received signal power and signal quality both reduced and resulted in lower data rates.

These representative works and many other related research efforts ideally need to get tested in real-world scenarios with autonomous UAVs to validate the main findings.  DTs such as AERPAW \cite{ae_1} are vital platforms for development and testing in a virtual environment before real-world deployment.
Experimenters can first access the DT resources online to start an experiment with provided sample experiments on the portal. Subsequently, those experiments are seamlessly moved to the outdoor AERPAW testbed that includes a 5-square-mile outdoor area with dedicated towers and drone testing facilities in North Carolina. Experiments can be done in urban and rural settings with UAV dynamic environments. Moreover, the DT supports replicating and refining experiments for a more accurate evaluation of UAV communications performance. 
Authors in \cite{ae_pre} \cite{gurses2024} showed the AERPAW DT’s ability for emulations in UAV communications and autonomy by testing various network configurations and fine-tuning network settings. 

Despite the advancements in simulation and DT technologies, there is a significant sim-to-real gap because of the inherent complexities of environmental variables and system dynamics. This motivates us to develop a simulation framework to bridge this gap and provide more accurate predictions for UAV communications in wireless networks.  Table~\ref{tab:comparison} shows a comparison of DT, simulation, and real-world characteristics for UAV communication to understand the gap. It is clear in Table~\ref{tab:comparison} that deployment in real-world environments is more critical and hence more advanced algorithm design is required to resemble the real-world scenario with the predicted scenarios of simulation frameworks and DTs. 

The integration of UAVs with digital twins (DTs) is becoming increasingly popular, enabling a wider range of practical use cases. For example, during natural or manmade disaster \cite{WON2023385} scenarios, DTs provide a platform for pre-deployment testing of UAV networks which helps robust connectivity and effective coordination in areas with limited infrastructure. Another attractive use case of UAVs within DT environments is their use as data mules where without any physical environments, UAVs can act as mobile data collectors and deliverers. Hence, they can bridge the gap between remote locations and central processing systems with sparse network connectivity, such as rural areas. Our proposed flexible simulation framework can accommodate these use cases with custom-programmed modifications, accelerating the subsequent development cycle with the DT and real-world testing.
\begin{table}[t!]
\centering
\caption{Comparison among DT, simulation abstractions for DT, and real-world characteristics for UAV communication.}
\begin{tabular}{|p{0.65in}|p{0.65in}|p{0.65in}|p{.8in}|}
\hline
\textbf{} & \textbf{DT} & \textbf{Simulation Abstractions} & \textbf{Real World} \\
\hline
\textbf{Protocols} & Full stack & Shannon capacity & Full stack with unexpected protocol delays\\
\hline \textbf{Wireless Software} & srsRAN, OAI, GNU radio & Matlab & Proprietary equipment\\
\hline \textbf{Drone Software} & AirSim, PX4 & Matlab & Onboard flight controllers\\
\hline \textbf{Propagation Effects} & FSPL, antenna effects & FSPL, antenna effects & Multipath, interference\\   
\hline \textbf{Hardware Impairments} & None & None & Hardware aging, noise\\
\hline \textbf{Runtime} & Long & Short & Real-time\\
\hline \textbf{Complexity} & High & Low & Very high\\
\hline \textbf{Calibration} & Complex & Flexible & Requires field testing\\
\hline
\end{tabular}%
\label{tab:comparison}
\end{table}

\section{System Model and Path Loss Calculation}\label{smpl}
\begin{figure}[t!]
\centerline{\includegraphics[width=0.5\textwidth]{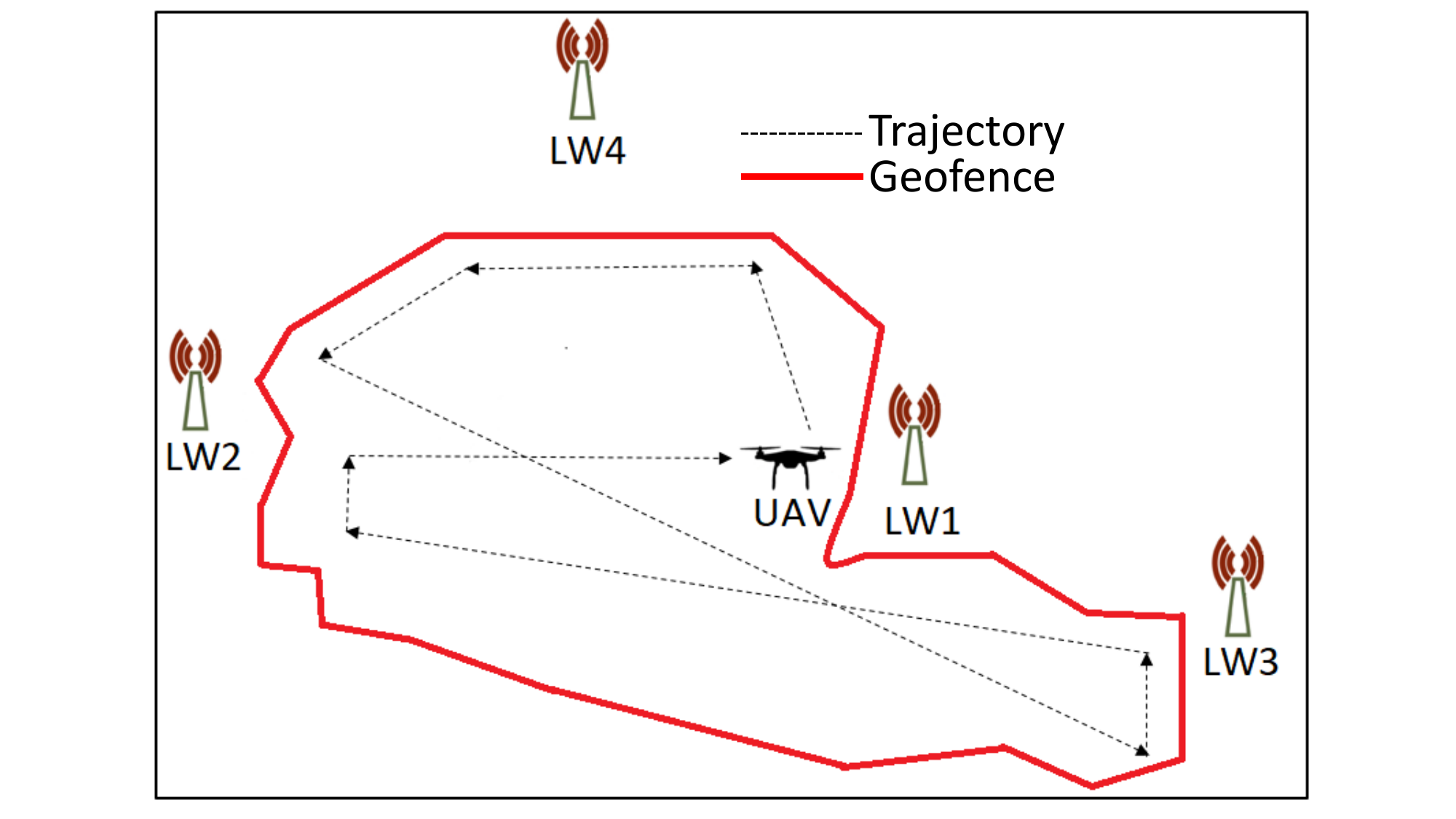}}
\caption{An example of a mission where a UAV flies to collect data from multiple towers.}
\label{system_model1}\vspace{-2mm}
\end{figure}

We consider a wireless network as shown in Fig.~\ref{system_model1}, where UAVs interact with a fixed wireless infrastructure while they fly. As an example, Fig.~\ref{system_model1} provides the towers and flight area from the NSF AERPAW  \cite{ae_1} testbed, which provides a realistic environment for evaluating UAV communication performance. Here, UAVs traverse predetermined trajectories within AERPAW's phase-1 geofence area ~\cite{ae_1}. 

\subsection{Path Loss Calculation}\label{plc}
Path loss is critical to model for UAV networks as it significantly affects signal propagation, link reliability, and overall network performance. To accurately estimate path loss, we need to calculate the distance between the UAV and the base station (BS). Hence, the two-dimensional distance $d$ 
 between two latitude-longitude points for UAV and the BS is given by the Haversine formula: 

\begin{equation}    
\begin{aligned}
d &= 2r\,\text{arcsin}\Bigg(\Bigg[\sin^2\left(\frac{\phi_{\rm b} - \phi_{\rm u}}{2}\right) \\
  &\quad + \cos(\phi_{\rm u})\cos(\phi_{\rm b})\sin^2\left(\frac{\lambda_{\rm b} - \lambda_{\rm u}}{2}\right)\Bigg]^{1/2}\Bigg)
\label{eq}
\end{aligned}
\end{equation}

where $\phi_{\rm u},~ \phi_{\rm b}$ are the latitude and $\lambda_{\rm u},~ \lambda_{\rm b}$ are the longitude of UAV and BS, respectively, in radian form, and $r$ is the radius of the earth. 
The three-dimensional distance $D$ between a UAV and a BS is given as: 
\begin{equation}
D=\sqrt{d^2+h^2}~, \quad  h=h_{\rm b}-h_{\rm u}~,
\end{equation}
where \( h_u \) and \( h_b \) are the altitude of the UAV and the BS in meters. Note that while other standardized channel models exist for UAVs, we initially use free space path loss (FSPL) \cite{fspl} with antenna effects to align DT results with simulation abstractions.
\begin{equation}
\begin{aligned}
\operatorname{FSPL} 
& =20 \log _{10}(D)+20 \log _{10}(f)-147.55~,
\end{aligned}
\end{equation}
where $D$ is the 3D distance between the UAV and the BS. 
The RSRP is calculated as:
\begin{equation}
\operatorname{RSRP} = P_{\rm tx} + G_{\rm tx} - \text{FSPL} + G_{\rm rx} + \delta_{\mathrm{LW_{i}}}~, \quad i=1,2,3,4~,\nonumber
\end{equation}
where $P_{\rm tx}$ is the transmit power of BS, 
$G_{\rm tx}$ and $G_{\rm rx}$ are the antenna gains of the BSs and UAV, respectively, and $\delta_{\text{LW}i}$ represents a calibration offset for the radios at  $\mathrm{LW}_{\rm 1}, \ \mathrm{LW}_{\rm 2}, \ \mathrm{LW}_{\rm 3}, \ \mathrm{LW}_{\rm 4}$.

Then, we calculate the receive power, $P_\text{rx}$ and the signal-to-noise ratio (SNR) \cite{sesiabook}, as follows: 
\begin{equation}
P_{\rm rx}=\mathrm{RSRP}~, 
\operatorname{SNR} =\frac{P_{\rm rx}}{N}~.
\end{equation}

Table \ref{tab:parameters} shows the parameters with values used in this experiment. We consider the maximum speed of the UAV to be $v_{\rm f}=5$~ m/s for a fixed trajectory and $v_{\rm a}=10$~m/s for an autonomous trajectory.

\begin{table}[t]
\centering
\caption{Parameter settings for experiments.}
\begin{tabular}{cc|cc}
\hline
\textbf{Parameters} & \textbf{Values} & \textbf{Parameters} & \textbf{Values} \\
\hline
$h_{\rm b}$ & $10$~m & $G_{\rm rx}$ & $2$~dBi \\
$h_{\rm u}$ & $30$~m & $G_{\rm tx}$ & $10$~dBi \\
$P_{\mathrm{tx}}$ & $10$~dBm & $v_{\rm f}$ & $0$-$5$~m/s \\
$P_{\mathrm{N}}$ & $-90$~dBm & $v_{\rm a}$ & $0$-$10$~m/s \\
\hline
\end{tabular}
\label{tab:parameters}
\end{table}

\section{Radio Software and Propagation Channel Abstraction}\label{rsoft}
Efficient UAV communication relies on accurate modeling of the radio environment. Here, we discuss the abstraction of the propagation channel and the integration of radio software to support UAV operations as in Algorithm~\ref{alg:log_report}. The radio software models signal strength, SNR, and other channel parameters, which are essential for trajectory planning and decision making. 
We wrote an event-driven software in Matlab so that every second the Algorithm~\ref{alg:log_report} is executed 
to measure all RSRPs based on the current location of the UAV. 
This function stores all the necessary information in several files for further post-processing of simulated data. In Algorithm~\ref{alg:log_report}, the distance from the UAV to a BS is calculated using the haversine formula. Then, FSPL, RSRP, and SNR are calculated. Finally, the data rate is calculated according to the SNR and channel quality indicator (CQI) mapping table in LTE
\cite{sesiabook}.  Table \ref{tab:symbols} shows the variable definitions used in this paper.

\begin{table}[t]
\centering
\caption{Variable definitions used in this paper.}
\begin{tabular}{cc|cc}
\hline
\textbf{Notation} & \textbf{Definition} & \textbf{Notation} & \textbf{Definition} \\
\hline

$d_{\rm min}$ & min. dist. of a BS from UAV & $t_{\rm f}$ & flight time\\

$d_{\rm i}$ & list of distance from all BS & $t_{\rm e}$ & elapsed time\\

$d_{\rm th}$ & min. thresh. dist. (100 m) & $d'$ & current dist.\\

$d_{\rm ms}$ & min. safe dist. (50 m) & $d''$ & previous dist.\\ 

$\Delta \lambda$ & longitude difference & $\alpha_{\rm i}$ & initial step size\\

$\Delta \phi$ & latitude difference &  $\alpha$ & new step size\\

$\Delta \alpha'$ & updated latitude step size & $\theta$ & rotation angle\\

$\Delta \alpha''$ & updated longitude step size & $\Theta$ & set of angles\\

$\mathcal{D}$ & three dimensional dist. & $\mathcal{L}_{\rm FS}$ & FSPL\\

$\delta_{\mathrm{LW}_{\rm i}}$ & fixed offset for indv. BSs & $\mathcal{R}$ & RSRP\\

$\eta_{\rm {s}}$ & spectral eff. (lookup table) & $\gamma$ & SNR \\





$C_{\rm \eta}$ & throughput from $\eta_{\rm s}$ 

\end{tabular}
\label{tab:symbols}
\end{table}

\begin{algorithm}[t]
    \caption{UAV log report generation}
    \label{alg:log_report}
    \begin{algorithmic}[1]
       \STATE \textbf{Input:} $P_{\rm tx}$,~$P_{\rm N}$,~$f$,~$\phi_{\rm u}$,~$\phi_{\rm b}$,~$\lambda_{\rm u},~ \lambda_{\rm b}$,~$h_{\rm u}$,~$h_{\rm b}$
        \STATE \textbf{Output:} $t$,~$\phi_{\rm u}$,~$\lambda_{\rm u}$,~$h_{\rm  u}$,~$\mathcal{R}$,~$P_{\rm rx}$,~$\gamma$,~$\mathcal{C}$
        \FOR{$k \gets 1$ \TO $4$}
            \STATE $\mathcal{D} \gets 
            \text{haversine}(\phi_{\rm u},~\phi_{\rm b},~\lambda_{\rm u},~ \lambda_{\rm b},~h_{\rm u},~h_{\rm b})$ 
            \STATE $\mathcal{L}_{\rm FS} \gets 20~\log~\mathcal{D} + 20~\log~f - 147.55$

            \STATE $\mathcal{R} \gets P_{\rm tx} + G_{\rm tx} - \mathcal{L}_{\rm FS} + G_{\rm rx} + \delta_{\mathrm{LW}_{\rm i}}$

            

            
            \STATE $P_{\rm rx} \gets \mathcal{R}$ 
            
            \STATE $\gamma \gets ~P_{\rm rx} - ~P_{\rm N}$
            



            \STATE $C_{\rm \eta} \gets 1.4 \times (\eta_{\rm s} \gets \gamma)$

            
            
            \STATE $t \gets \text{datestr}(\text{now})$                      
            \STATE $\phi_{\rm u} \gets \phi_{\rm u}~,~\lambda_{\rm u} \gets ~\lambda_{\rm u}~,~h_{\rm u} \gets h_{\rm u}$
        \ENDFOR        
    \end{algorithmic}
\end{algorithm}

\section{Vehicle Software and Mission Plan Abstraction}\label{trjlog}
Based on the radio software's insights, the vehicle software abstracts the mission planning process and eases the autonomous and fixed trajectory operations. In this Section, we discuss the algorithms and decision-making processes along with the consideration of geofencing, trajectory optimization, and adaptability to real-time challenges. One of the major goals of development in an autonomous UAV network DT is to optimize the trajectories of one or more UAVs. In particular, UAVs can follow a fixed trajectory with predetermined waypoints or autonomously optimize their trajectories in real time based on environmental conditions and mission requirements. 

\subsection{Autonomous Trajectory}\label{atr} 
In this Section, we propose an algorithm for autonomous trajectory based on Euclidean distance between the UAV and all the BSs. A specific example use case is considered where the UAV autonomously navigates to visit the closest point to each BS at the geofence boundary. 
This scenario represents a practical application of the proposed framework but is only one of many potential use cases for autonomous flight planning, such as for data mules~\cite{data_mule}. The strategy is that the distance from the UAV to all the BSs is calculated using the Euclidean approach and then the UAV tries to reach the nearest BS and also maintains the geofence. UAV will not move out of the geofence and when it reaches near the geofence, it will change its orientation to find a new path to reach that BS. Moreover, the UAV can move slower and faster depending on its distance from a BS. If it is near a BS, its speed slows down gradually. Then, it changes its turn towards the next BS. This process continues until the UAV visits all the BSs or the flight time exceeds. Finally, the UAV returns to its launch position and lands. Autonomous trajectory has been illustrated in Algorithm \ref{alg:uav_navigation}. Fig ~\ref{merge}(a) shows the drone's autonomous trajectory after completing its mission. Here, depending on the current position of the UAV from the BS, the UAV chooses first LW1 to reach, then LW2, then LW4, and finally LW3. Then, the UAV returns to its launch position and lands.

\subsection{Fixed Trajectory}\label{ft}
\begin{figure}[t]
\centerline{\includegraphics[width=0.52\textwidth, trim=2cm 2.75cm 2cm 3.5cm, clip]{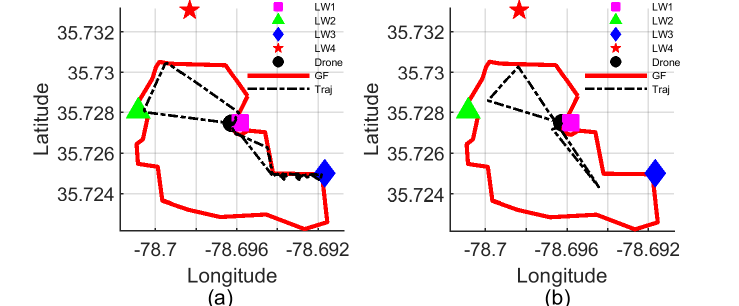}}
\caption{Simulations for UAV flight missions considering
 (a) autonomous and (b) fixed trajectory.}
\label{merge}
\end{figure}

There are some use cases where a fixed or predetermined trajectory is helpful to ensure safe and efficient navigation. An example use case for a fixed trajectory is the area where UAV operations are restricted to pre-approved flight paths to avoid restricted zones such as near airports or military installations. For the fixed trajectory, we first compute the number of steps for ascending and descending and the steps that the UAV travels through two intermediate waypoints. These steps are calculated by dividing the distance between two fixed points by the UAV's speed. We then calculate the UAV positions for those fixed steps using the linespace function in Matlab, and the UAV moves toward its target following these predefined flight paths.
The proposed system has been implemented in the Matlab environment as shown in Fig. ~\ref{merge}(b). We modified the 'kml2struct.m' file to extract geofence data of \cite{ae_1} in the correct format in our experiment. It is an animated experiment and at the same time, reports are generated similarly to what is available in the DT environment using the mission control software. In the example of Fig.~\ref{merge}(b) the UAV first flies vertically to $30$~m and travels horizontally following the trajectory. We use a fixed trajectory to generate a similar RSRP between the AERPAW emulation and Matlab simulation to ensure our developed framework works as intended. A comparison of the emulation and simulation in terms of RSRP measurements is shown in Section~\ref{V(A)}.

\begin{algorithm}[t]
    \caption{Autonomous UAV flight navigation}
    \label{alg:uav_navigation}
    \begin{algorithmic}[1]
    

    \STATE \textbf{Input:} \( t_{\mathrm{f}},~ t_{\mathrm{e}},~ \phi_{\mathrm{u}},~ \lambda_{\mathrm{u}},~ \phi_{\mathrm{b}},~ \lambda_{\mathrm{b}},~ d_{\mathrm{th}},~ \alpha_{\mathrm{i}},~ v_{\mathrm{i}},~ \Theta \)

    \STATE \textbf{Output:} Navigation to the nearest BS
    
   \WHILE{not empty~(\( \phi_{\text{b}} \))} 
        \IF{$t_{\rm e} \geq t_{\rm f}$}
            \STATE \textbf{break}
        \ENDIF 

        \STATE $d_{\rm i}~ \gets 
        {\rm pdist2}~(\phi_{\rm u},~\phi_{\rm b},~\lambda_{\rm u},~\lambda_{\rm b})$

        \STATE $d_{\rm min}~,~ i \gets \min~(d_{\rm i})$

        \STATE $\phi'_{\rm b} \gets \phi_{\rm b}(i)~, ~\lambda' _{\rm b} \gets \lambda_{\rm b}(i)$        

        \STATE $\Delta \phi \gets \phi'_{\rm b} - \phi _{\rm u}~,~ \Delta \lambda \gets \lambda'_{\rm b} - \lambda _{\rm u}$
        \STATE $d' \gets \sqrt{{\Delta \phi}^2 + {\Delta \lambda}^2}$ 


        \IF{$d_{\rm min} > d_{\rm th}$}
            \STATE $v_{\rm i} \gets v_{\rm i}$
        \ELSE
            \STATE $v_{\rm i} \gets \frac{v_{\rm i}}{2}$
        \ENDIF
        \STATE $\alpha \gets ~v~\alpha _{\rm i}~,~v\gets v_{\rm i}~$
        

        \IF{$d' > d_{\rm ms}$}
            \STATE $d'' \gets \infty$

            \WHILE{$d' < d''$}
                \STATE $\Delta {\alpha'} \gets \left(\frac{\Delta \phi}{d'}\right) \alpha~,~ \Delta {\alpha''} \gets \left(\frac{\Delta \lambda}{d'}\right) \alpha$

                \STATE $\phi_{\rm u} \gets \phi_{\rm u} + \Delta {\alpha'}~,~\lambda_{\rm u} \gets \lambda_{\rm u} + \Delta {\alpha''}$

                \IF{is\_in\_geofence(${\phi _{\rm u}}, {\lambda _{\rm u}}$)}
                    \STATE $\text{Update drone position with} ~{\phi _{\rm u}}, {\lambda _{\rm u}}, {h _{\rm u}}$
                    \STATE $d'' \gets d'~$
                    \STATE $\Delta \phi \gets \phi'_{\rm b} - \phi _{\rm u}~,~ \Delta \lambda \gets \lambda'_{\rm b} - \lambda _{\rm u}$
                    \STATE $d' \gets \sqrt{{\Delta \phi}^2 + {\Delta \lambda}^2}$
                    

                    \IF{$d' \geq d''$}
                        \IF{$d' < d_{\rm th}$}
                            \STATE Mark a BS as visited 
                            \STATE \textbf{break}
                        \ENDIF
                    \ENDIF
                \ELSE

                    
                    \STATE $\theta \in \Theta~,~\Delta\phi' \gets \Delta\phi$ 
                    \STATE $\Delta \phi \gets \Delta \phi'~ \cos\theta - \Delta \lambda ~\sin\theta$
                    \STATE $\Delta \lambda \gets \Delta\phi'~\sin\theta+\Delta\lambda~~\cos\theta$
                    
                    \STATE $d' \gets \sqrt{{\Delta \phi}^2 + {\Delta \lambda}^2}$
                \ENDIF
            \ENDWHILE
        \ELSE
            \STATE Mark a BS as visited
            \STATE Rotate UAV towards the next BS
        \ENDIF
    \ENDWHILE
    \end{algorithmic}
\end{algorithm}

\section{Result and Discussion}\label{rd}
To compare the results from AERPAW's DT and our Matlab simulations including radio/vehicle software abstractions, we carried out experiments with a fixed trajectory for UAVs in both environments and only the autonomous trajectory in simulation.
We set the downlink frequency, $f = 3410$~MHz, similar to the srsRAN LTE SISO experiment \cite{ae_1}. In Section \ref{V(A)}, we compare the RSRP for fixed trajectory in emulation and simulation and also show the RSRP for autonomous trajectory in simulation. In Section \ref{BB}, we discuss the theoretical throughput for autonomous trajectory in simulation.

\begin{figure*}[t]
\centerline{\includegraphics[width=\textwidth,trim=1cm .7cm .9cm .9cm, clip]
{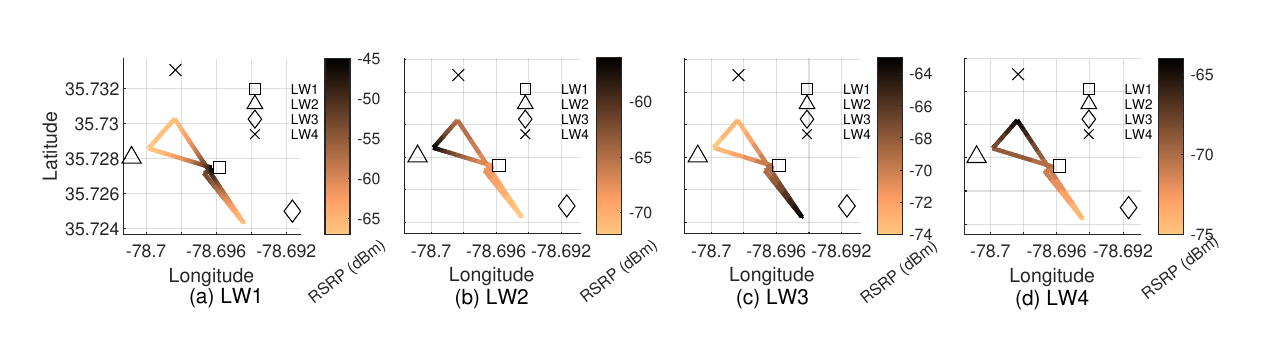}}
\caption{RSRP measurement with respect to (a) LW1, (b) LW2, (c) LW3, and (d) LW4 for fixed trajectory in emulation.}
\label{rsrp}
\vspace{-2mm}
\end{figure*}

\begin{figure*}[t]
\centerline{\includegraphics[width=\textwidth,trim=1cm .7cm .9cm .9cm, clip]{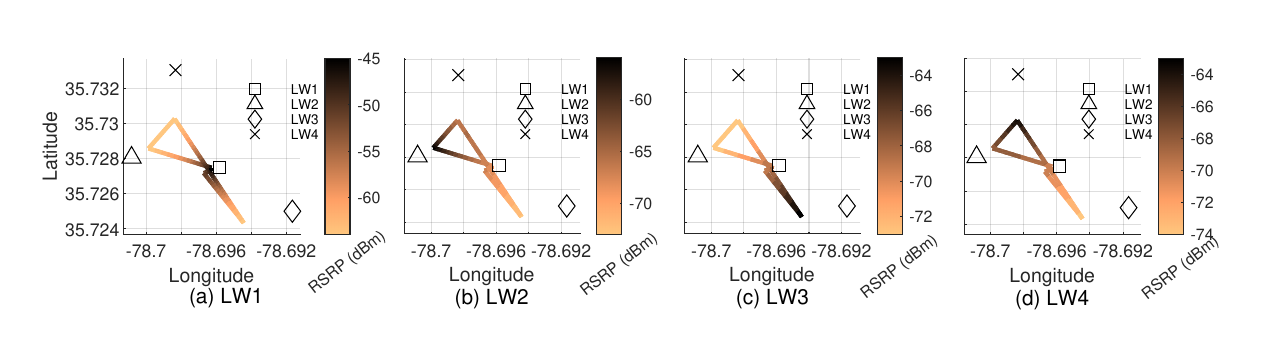}}
\caption{RSRP measurement with respect to (a) LW1, (b) LW2, (c) LW3, and (d) LW4 for fixed trajectory in simulation.}
\label{rsrp2}
\end{figure*}

\begin{figure*}[t]
\centerline{\includegraphics[width=\textwidth,trim=1cm .6cm .9cm .9cm, clip]
{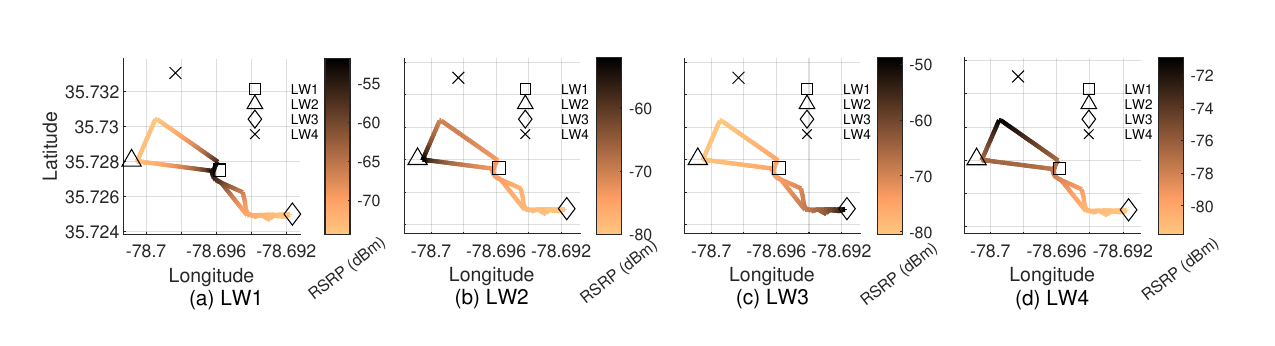}}
\caption{RSRP measurement with respect to (a) LW1, (b) LW2, (c) LW3, and (d) LW4 for autonomous trajectory in simulation.}
\label{rsrp3}
\vspace{-2mm}
\end{figure*}

\subsection{RSRP for Individual BSs}\label{V(A)}

In this section, we compare the simulation and emulation reports for fixed trajectory in Fig. \ref{rsrp} and Fig. \ref{rsrp2}. We also show the RSRP measurement for the autonomous trajectory within a geofence.
Fig.~\ref{rsrp} and Fig.~\ref{rsrp2} show the RSRP measurement by UAV for individual BSs in emulation and simulation, respectively. When the UAV is closer to a BS, signal strength tends to be stronger. For example, we observe a strong signal strength in Fig.~\ref{rsrp}(a) and Fig.~\ref{rsrp2}(a) because the UAV is close to the LW1 compared to other towers. Hence, we observe a high similarity in signal strength between simulation and emulation.

Fig. \ref{rsrp3} shows the RSRP measurement by UAV to individual BSs for autonomous trajectory in simulation. We see a strong RSRP near all BSs except LW4 because UAV goes near LW1, LW2, and LW3, but LW4 is far away and the UAV is restricted to geofence to reach there and so it cannot go near LW4. Hence, we see weaker signal strengths for LW4.

\begin{figure}[t]
\centerline{\includegraphics[width=0.5\textwidth]{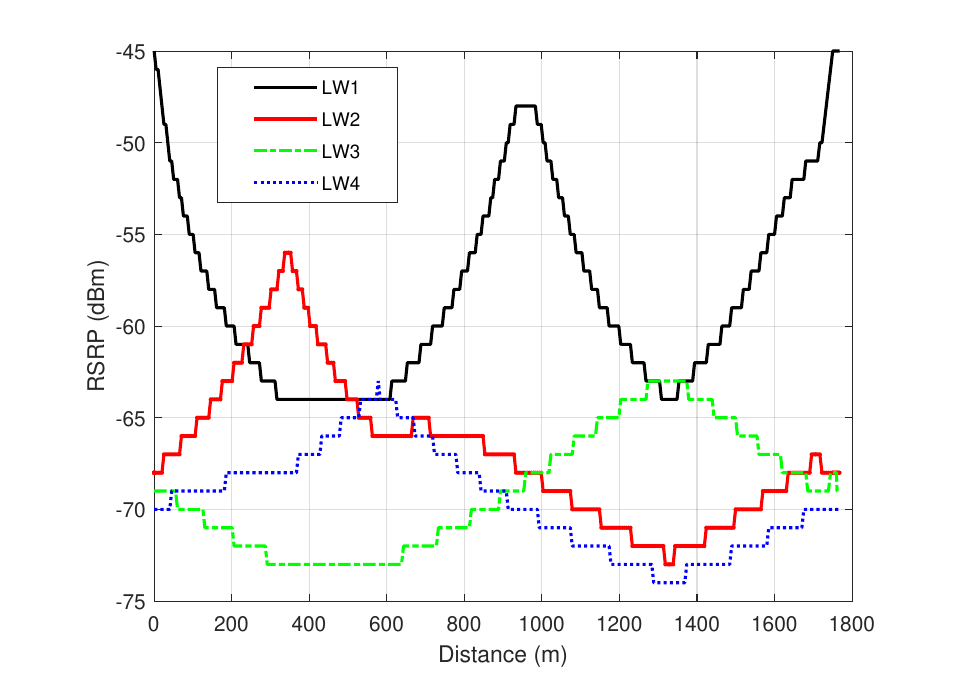}}
\caption{RSRP with distance measured by UAV from multiple BSs for fixed trajectory in simulation.}
\label{fix_sim_rsrp_distance}
\vspace{-5mm}
\end{figure}

\begin{figure}[t]
\centerline{\includegraphics[width=0.5\textwidth]{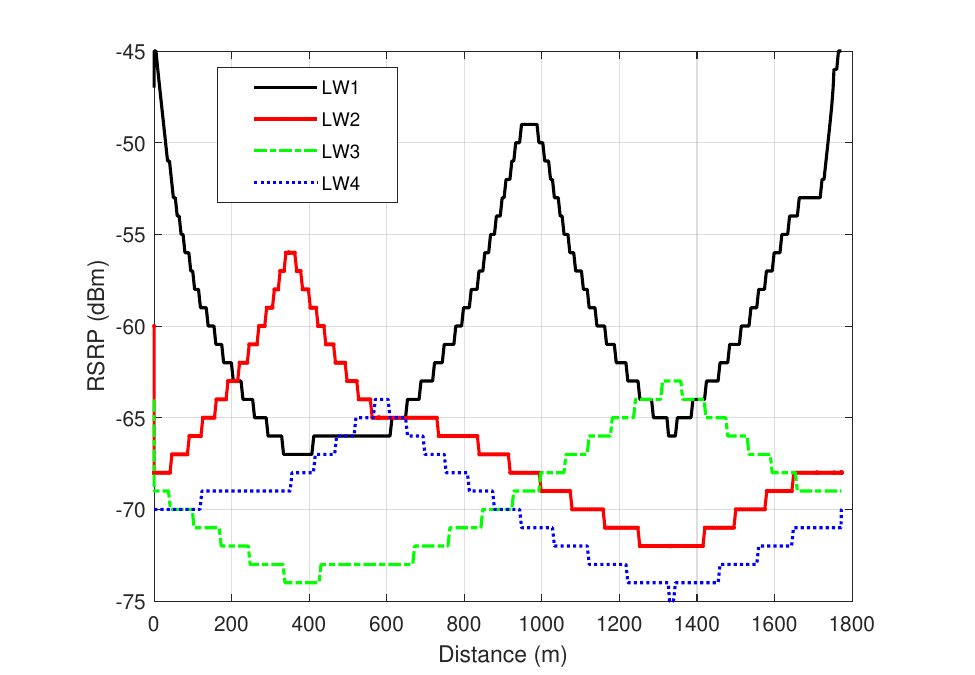}}
\caption{RSRP with distance measured by UAV from multiple BSs for fixed trajectory in emulation (using srsRAN).}
\label{fix_emu_rsrp_distance}
\vspace{-5mm}
\end{figure}

Fig. \ref{fix_sim_rsrp_distance} and Fig.~\ref{fix_emu_rsrp_distance} show the RSRP measurement with distance observed by UAV from all the BSs for fixed trajectory in simulation and emulation, respectively. When the UAV is closer to a BS, its RSRP becomes stronger. UAV first goes to LW1 from its launch position. After reaching there, it goes to LW2, LW4, and LW3, as shown in Fig.~\ref{merge}(b). 
From there it returns to its launch position again to land. Throughout the flight mission, the distance from the UAV to the LW1 is typically small compared to other towers, which is why RSRP at LW1 is higher. Here, from Fig.~\ref{fix_sim_rsrp_distance} and Fig.~\ref{fix_emu_rsrp_distance}, we find approximately similar RSRP with the distance between the simulation and emulation environments.

\subsection{Throughput}\label{BB}
\begin{figure}[t]
\centerline{\includegraphics[width=0.5\textwidth]{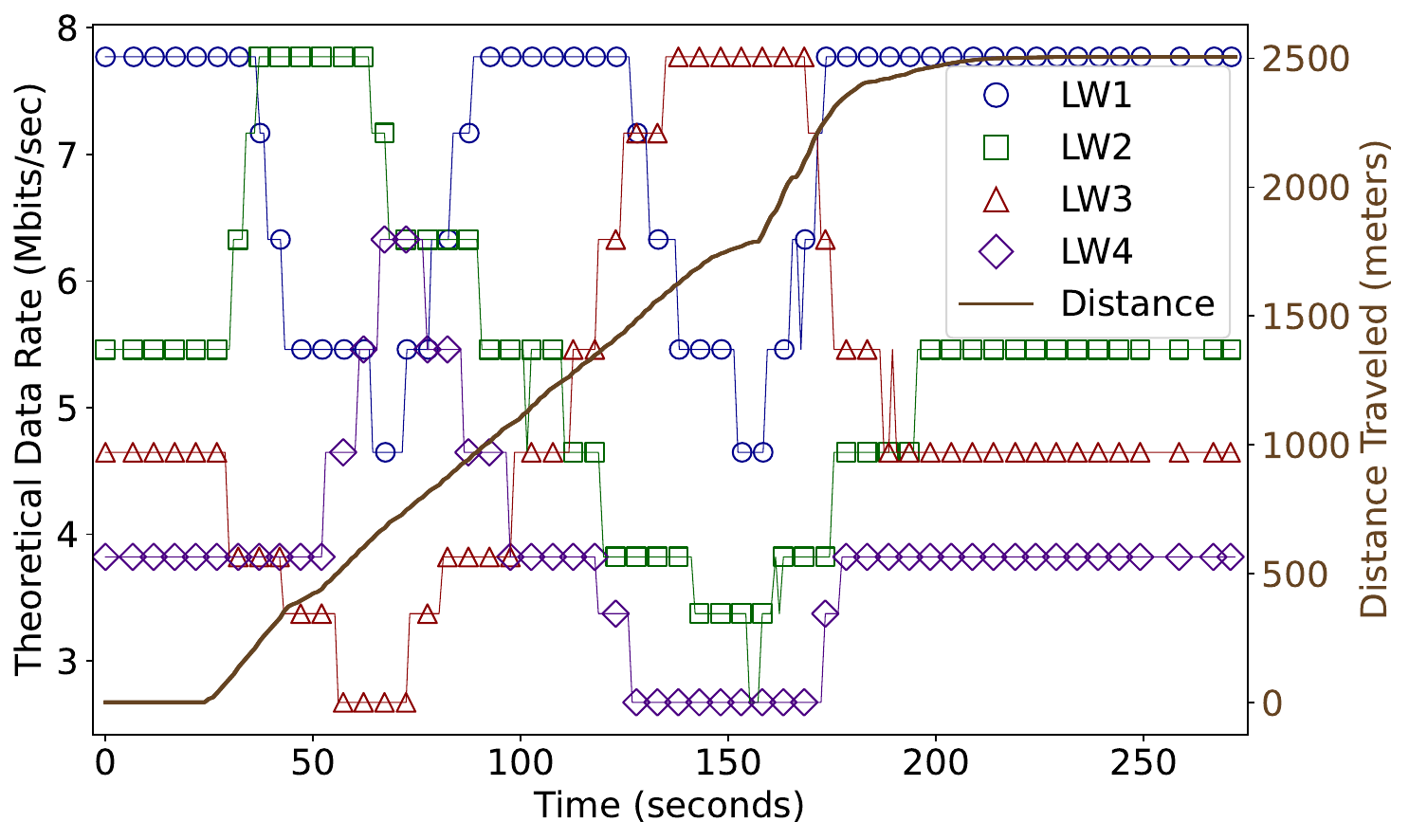}}
\caption{Simulated throughput for autonomous trajectory.}
\label{fig:th_at}
\vspace{-3mm}
\end{figure}

Due to good signal strength, higher throughput is expected when the UAV is closer to a BS, and when it moves far away from a BS, its data rate decreases because of the lower signal strength.
Fig. \ref{fig:th_at} shows the throughput for the autonomous trajectory with the changes of time and distance calculated in our developed simulated framework. 
The simulated throughput has been calculated using the SNR and CQI mapping table, i.e., $C_{\rm \eta}$ mentioned in Algorithm \ref{alg:log_report}. When the UAV remains within a good coverage area with an SNR value of $22.7$~dB or higher, CQI = $15$ and $64$~QAM is applied \cite{sesiabook}. Having $1.4$~MHz bandwidth, a maximum throughput of $7.77$~Mbits/sec is observed.
In Fig. \ref{fig:th_at}, between 1-40 seconds, the UAV is closer to LW1 compared to other towers, and hence at that time we see the maximum data download rate (DDR) at 7.77 Mbits/sec. Between 40 and 55 seconds, the UAV is closer to LW2 and we see the DDR at LW2 is higher. Now, UAV goes towards the LW4 but LW4 is far away from the geofence compared to the other towers, and hence the DDR does not reach the highest rate. Then, the UAV moves towards the LW3, but it can download from LW1 at the maximum rate between 70 and 120 seconds. When the UAV goes near LW3 between 125 and 170 seconds, it can download at a maximum rate from LW3 but the download rate at LW4 is minimum because LW4 is far away at that interval. Hence, the signal strength of the LW4 from the UAV's current position at that time is much lower, which results in lower DDR. UAV then moves forward to its launch position and hence the DDR from LW1 is higher due to the good signal strength and closer distance to the UAV.

Additional work has been in progress to generate throughput results using the emulation environment and srsRAN, to compare with the results similar to the scenario in Fig. \ref{fig:th_at}. Different key performance indicators from those scenarios will be reported in our future work for various different topologies and network configurations. 




\section{Conclusion}\label{con}
This paper presents a framework for simulation abstractions for modeling radio and vehicle behavior in an autonomous UAV network DT environment. The simulation abstractions allow faster development and optimization of artificial intelligence and machine learning algorithms without running a full-stack protocol software in the DT. On the other hand, the proposed simulation framework allows quickly deploying the developed software to the DT and then to the testbed environment. The developed software is available at~\cite{uavsimframework} for broader use by the research community. 
Our future work includes incorporating additional features such as trajectory optimization, ray tracing, airframe shadowing, interference effects, and dynamic changes due to Doppler and UAV mobility.

\bibliographystyle{IEEEtran}
\bibliography{main}

\begin{thebibliography}{10}
\providecommand{\url}[1]{#1}
\csname url@samestyle\endcsname
\providecommand{\newblock}{\relax}
\providecommand{\bibinfo}[2]{#2}
\providecommand{\BIBentrySTDinterwordspacing}{\spaceskip=0pt\relax}
\providecommand{\BIBentryALTinterwordstretchfactor}{4}
\providecommand{\BIBentryALTinterwordspacing}{\spaceskip=\fontdimen2\font plus
\BIBentryALTinterwordstretchfactor\fontdimen3\font minus \fontdimen4\font\relax}
\providecommand{\BIBforeignlanguage}[2]{{%
\expandafter\ifx\csname l@#1\endcsname\relax
\typeout{** WARNING: IEEEtran.bst: No hyphenation pattern has been}%
\typeout{** loaded for the language `#1'. Using the pattern for}%
\typeout{** the default language instead.}%
\else
\language=\csname l@#1\endcsname
\fi
#2}}
\providecommand{\BIBdecl}{\relax}
\BIBdecl

\bibitem{in_1}
R.~Shahzadi, M.~Ali, H.~Z. Khan, and M.~Naeem, ``{UAV} assisted {5G} and beyond wireless networks: A survey,'' \emph{Journal of Network and Computer Applications}, vol. 189, p. 103114, 2021.

\bibitem{ae_1}
\BIBentryALTinterwordspacing
``Aerial experimentation and research platform for advanced wireless {(AERPAW)}.'' [Online]. Available: \url{https://aerpaw.org/}
\BIBentrySTDinterwordspacing

\bibitem{dt}
D.~Villa, M.~Tehrani-Moayyed, C.~P. Robinson, L.~Bonati, P.~Johari, M.~Polese, and T.~Melodia, ``Colosseum as a digital twin: Bridging real-world experimentation and wireless network emulation,'' \emph{IEEE Trans. Mobile Compu.}, vol.~23, no.~10, pp. 9150--9166, 2024.

\bibitem{rw_1}
M.~Mozaffari, W.~Saad, M.~Bennis, Y.-H. Nam, and M.~Debbah, ``A tutorial on {UAVs} for wireless networks: Applications, challenges, and open problems,'' \emph{IEEE Commun. Surveys \& Tut.}, vol.~21, no.~3, pp. 2334--2360, 2019.

\bibitem{max_thr}
L.~Xie, J.~Xu, and R.~Zhang, ``Throughput maximization for {UAV}-enabled wireless powered communication networks,'' \emph{IEEE Int. of Things J.}, vol.~6, no.~2, pp. 1690--1703, 2019.

\bibitem{lte_alt}
M.~A. Zulkifley, M.~Behjati, R.~Nordin, and M.~S. Zakaria, ``Mobile network performance and technical feasibility of {LTE}-powered unmanned aerial vehicle,'' \emph{Sensors}, vol.~21, no.~8, p. 2848, 4 2021.

\bibitem{ae_pre}
A.~Panicker, O.~Ozdemir, M.~L. Sichitiu, I.~Guvenc, R.~Dutta, V.~Marojevic, and B.~Floyd, ``{AERPAW} emulation overview and preliminary performance evaluation,'' \emph{Els. Comp. Netw.}, vol. 194, p. 108083, 2021.

\bibitem{gurses2024}
A.~Gürses, G.~Reddy, S.~Masrur, O.~Özdemir, I.~Güvenç, M.~Sichitiu, A.~Şahin, A.~Alkhateeb, M.~Mushi, and R.~Dutta, ``Digital twins and testbeds for supporting ai research with autonomous vehicle networks,'' \emph{IEEE Comm. Mag.}, 2024.

\bibitem{WON2023385}
J.~Won, D.-Y. Kim, Y.-I. Park, and J.-W. Lee, ``A survey on uav placement and trajectory optimization in communication networks: From the perspective of air-to-ground channel models,'' \emph{ICT Express}, vol.~9, no.~3, pp. 385--397, 2023.

\bibitem{fspl}
\BIBentryALTinterwordspacing
``Free space path loss.'' [Online]. Available: \url{https://en.wikipedia.org/wiki/Free-space_path_loss}
\BIBentrySTDinterwordspacing

\bibitem{sesiabook}
S.~Sesia, I.~Toufik, and M.~Baker, \emph{{LTE - The UMTS Long Term Evolution: From Theory to Practice}}, 2nd~ed.\hskip 1em plus 0.5em minus 0.4em\relax Wiley, 2011.

\bibitem{data_mule}
D.~Palma, A.~Zolich, Y.~Jiang, and T.~A. Johansen, ``Unmanned aerial vehicles as data mules: An experimental assessment,'' \emph{IEEE Access}, vol.~5, pp. 24\,716--24\,726, 2017.

\bibitem{uavsimframework}
\BIBentryALTinterwordspacing
``Uavflexsimframework,'' 2025. [Online]. Available: \url{https://github.com/mhossenece/UAVFlexSimFramework.git}
\BIBentrySTDinterwordspacing

\end{thebibliography}

\end{document}